\documentclass[12pt,preprint]{aastex}

\newcommand{\etal}{et~al.\ }

\newcommand{\cmsq}{\hbox{cm$^{-2}$}}

\newcommand{\lumin}{\hbox{ergs~s$^{-1}$}}

\newcommand{\nh}{\hbox{${N}_{\rm H}$}}
\newcommand{\h}{$h_{70}^{-1}$}

\newcommand{\chandra}{{\emph{Chandra}}}

\newcommand{\xmm}{\emph{XMM-Newton}}

\newcommand{\rosat}{\emph{ROSAT}}

\begin{document}

\def\sarc{$^{\prime\prime}\!\!.$}
\def\arcsec{$^{\prime\prime}$}
\def\arcmin{$^{\prime}$}
\def\degr{$^{\circ}$}
\def\seco{$^{\rm s}\!\!.$}
\def\ls{\lower 2pt \hbox{$\;\scriptscriptstyle \buildrel<\over\sim\;$}} 
\def\gs{\lower 2pt \hbox{$\;\scriptscriptstyle \buildrel>\over\sim\;$}} 
 
\title{The X-ray Properties of Optically-Selected Galaxy Clusters}

\author{Xinyu Dai, Christopher S. Kochanek, and Nicholas D. Morgan}

\altaffiltext{1}{Department of Astronomy,
 The Ohio State University, Columbus, OH 43210,
 xinyu, ckochanek, nmorgan@astronomy.ohio-state.edu}

\begin{abstract}
We stacked the X-ray data from the \rosat\ All Sky Survey for over 4,000 clusters selected from the 2MASS 
catalog and divided into five richness classes.  We detected excess X-ray emission over background at the 
center of the stacked images in all five richness bins.  The interrelationships between the mass,
X-ray temperature and X-ray luminosity of the stacked clusters agree well with those derived from
catalogs of X-ray clusters.  Poisson variance in the number of galaxies occupying halos of a given 
mass leads to significant differences between the average richness at fixed mass and the average mass 
at fixed richness that we can model relatively easily using a simple model of the halo occupation
distribution.  These statistical effects probably explain recent results in which optically-selected
clusters lie on the same X-ray luminosity-temperature relations as local clusters but have lower
optical richnesses than observed for local clusters with the same X-ray properties.  When we
further binned the clusters by redshift, we did not find significant redshift-dependent biases in the sense 
that the X-ray luminosities for massive clusters of fixed optical richness show little dependence on redshift
beyond that expected from the effects of Poisson fluctuations.  
Our results demonstrate that stacking of RASS data from optically selected clusters can be a powerful test for
biases in cluster selection algorithms.  
\end{abstract}


\section{Introduction}

The primary objective of many new extragalactic surveys is to determine the equation of state of
the dark energy.  One approach is to determine the evolution of the cluster mass function with redshift, 
which depends on dark energy through the growth factor and the volume element (e.g., Haiman, Mohr \& Holder 2001; 
Huterer \& Turner 2001; Podariu \& Ratra 2001; Kneissl et al. 2001; Newman et al. 2002; Levine, Schulz, \& White 2002; 
Majumdar \& Mohr 2003; Hu 2003).  Clusters are identified using some observable
proxy for their mass over a broad range of redshifts and the proxy must then be calibrated to
compare the results to the theoretically predicted mass functions.  Possible proxies and identification
methods are the optical richness of the cluster (e.g., Abell 1958; Huchra \& Geller 1982; Dalton et al. 1992; 
Postman et al. 1996, 2002; Zaritsky et al. 1997; Gonzalez et al. 2001; Bahcall et al. 2003; Gal et al. 2003; 
Gladders \& Yee 2005), the X-ray luminosity or temperature (e.g., Gioia et al. 1990; Henry et al. 1992; Rosati et al. 1995,1998; 
Ebeling et al. 1998, 2000; B\"ohringer et al. 2000, 2004; Bauer et al. 2002; Giacconi et al. 2002), the Sunyaev-Zeldovich decrement 
(e.g., Carlstrom et al. 2000; LaRoque et al. 2003) and the weak lensing shear 
(e.g., Wittman et al. 2001; Dahle et al. 2002; Schirmer et al. 2003).   Examples of large scale surveys planning
on using these methods are the Dark Energy Survey (DES\footnote{http://www.darkenergysurvey.org/}
using optical richness, Sunyaev-Zeldovich and weak lensing), the Large Synoptic Survey Telescope 
(LSST\footnote{http://www.lsst.org/} optical richness and weak lensing) and the Supernova / Acceleration Probe 
(SNAP\footnote{http://snap.lbl.gov/} optical richness and weak lensing).  This is in addition to smaller
scale surveys based on optical richness such as the SDSS cluster surveys (Goto et al. 2002; Bahcall et al. 2003)  and the
Red Cluster Sequence Survey (RCS, Gladders \& Yee 2005). 

A persistent worry about optically selected cluster samples is that chance projections of foreground
and background galaxies significantly bias the resulting catalogs.  The problem can range from false
positives, detections of non-existent clusters, to a richness bias in which chance projections 
lead to overestimates of the cluster richness.  The problems should be more severe for higher
redshift and lower richness clusters because the effects of a chance projection increase as the
number of detectable galaxies in the cluster diminishes.  One approach to checking the extent
to which a cluster finding algorithm is affected by these issues is to generate mock galaxy 
catalogs matching the actual data as closely as possible, find clusters in the mock catalogs
and then compare the output and input catalogs (e.g., Kochanek et al. 2003; Miller et al. 2005).  In
general these comparisons have found only modest biases.  A second approach is to compare the
mass estimates based on the optical richness to those from another method.  At present
this has meant examining either the X-ray properties (e.g., Bahcall 1977; Donahue et al. 2002; 
Kochanek et al. 2003;
Popesso et al. 2004) or the weak lensing masses (e.g., SDSS/RCS) of the optically-selected clusters.  

A basic problem of most existing tests of optically-selected cluster catalogs using X-ray data is
that the comparisons are made to catalogs of X-ray clusters rather than through X-ray observations of
the optically-selected clusters.  While this provides a simple means of calibrating the relationship
between optical and X-ray properties, it is not an optimal approach to searching for biases
in the optical selection methods.  The problems could be more severe because several recent
studies (e.g., Bower et al.\ 1997; Lubin, Mulchaey \& Postman 2004; Gilbank et al. 2004) have found offsets between the X-ray temperature-optical richness relations
of local and higher redshift optically-selected clusters even though the clusters lie on
the same X-ray temperature-luminosity relations.    

In this paper we address these questions by measuring the
mean X-ray properties of a sample of galaxy clusters selected from the 2MASS survey 
(Kochanek et al. 2003, 2006 in preparation) by averaging (``stacking") the \rosat\ All-Sky Survey (RASS, Voges et al. 1999)
data for the clusters as a function of richness and redshift. We outline the cluster
catalog and our procedures for analyzing the RASS data in \S2.  In \S3 we discuss
the theoretical differences between measuring the mean properties of clusters 
at fixed richness rather than fixed mass.  In \S4 we present our results for
the correlations between the optical richness, X-ray properties and cluster mass
estimates, as well as the dependence of the results on cluster redshift.  
We summarize our results
in \S5.  We assume that $H_0 = 70~\rm{km~s^{-1}~Mpc^{-1}}$, $\Omega_{\rm m} = 0.3$, 
and $\Omega_{\Lambda}= 0.7$ throughout the paper.

\begin{figure}
\epsscale{1}
\plotone{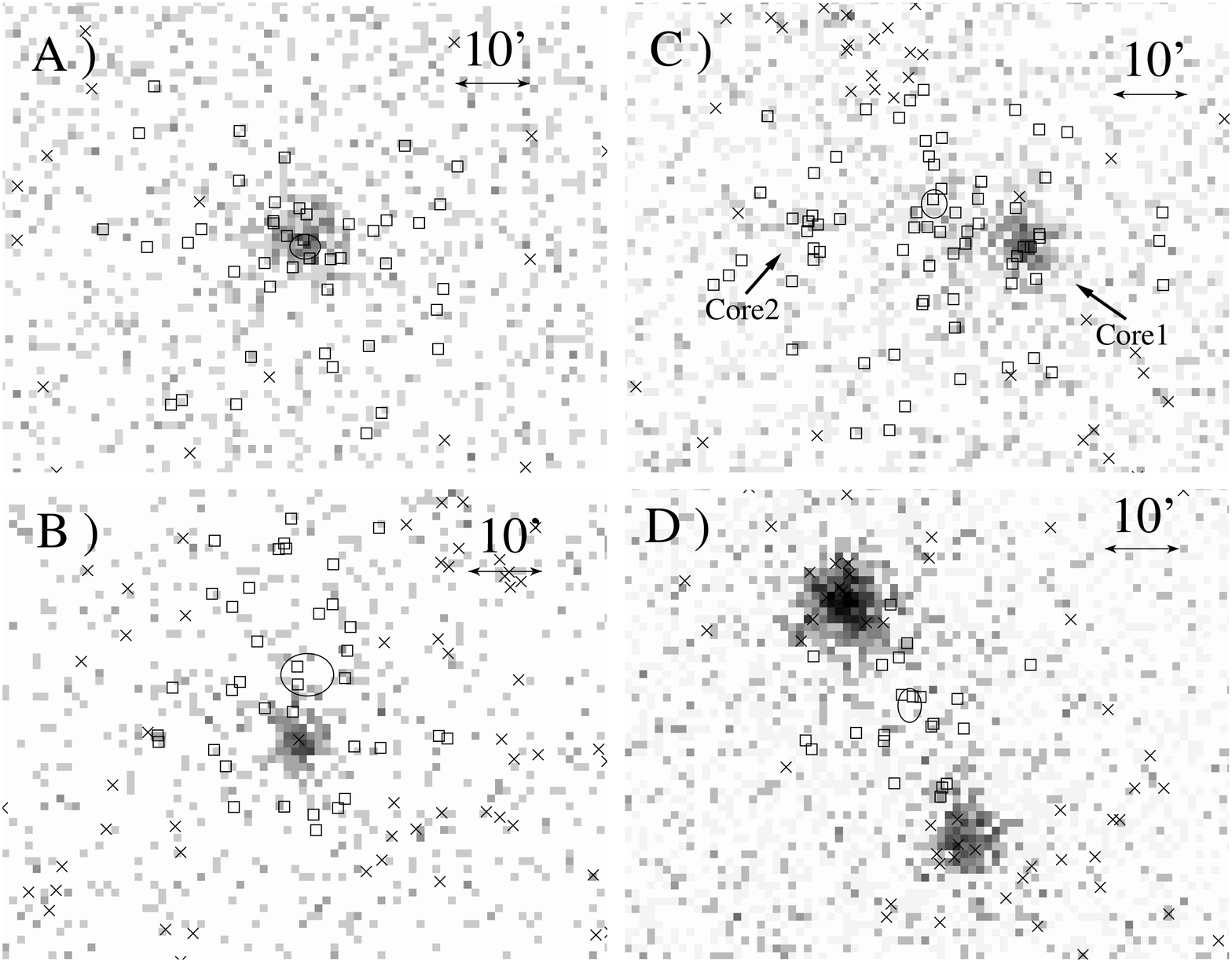}
\caption{RASS images of four clusters over-plotted with probable cluster members (squares, $> 50\%$ likelihood of membership as determined from the cluster finding algorithm in Kochanek et al. 2003) and field galaxies 
(X symbols, $< 50\%$ likelihood) along with the error ellipse for the cluster centroids provided by optical cluster detection algorithm.  
Panel A shows the typical cluster, where the optical position is roughly consistent with the X-ray position.  Panel B shows an example 
where the existence of a cluster is correctly inferred in the optical catalog, but the position is seriously in error.  Panels C and D 
show two examples where the optical cluster is confused by the presence of multiple (or merging) clusters.\label{fig:oxplot}}
\end{figure}

\begin{figure}
\epsscale{1}
\plotone{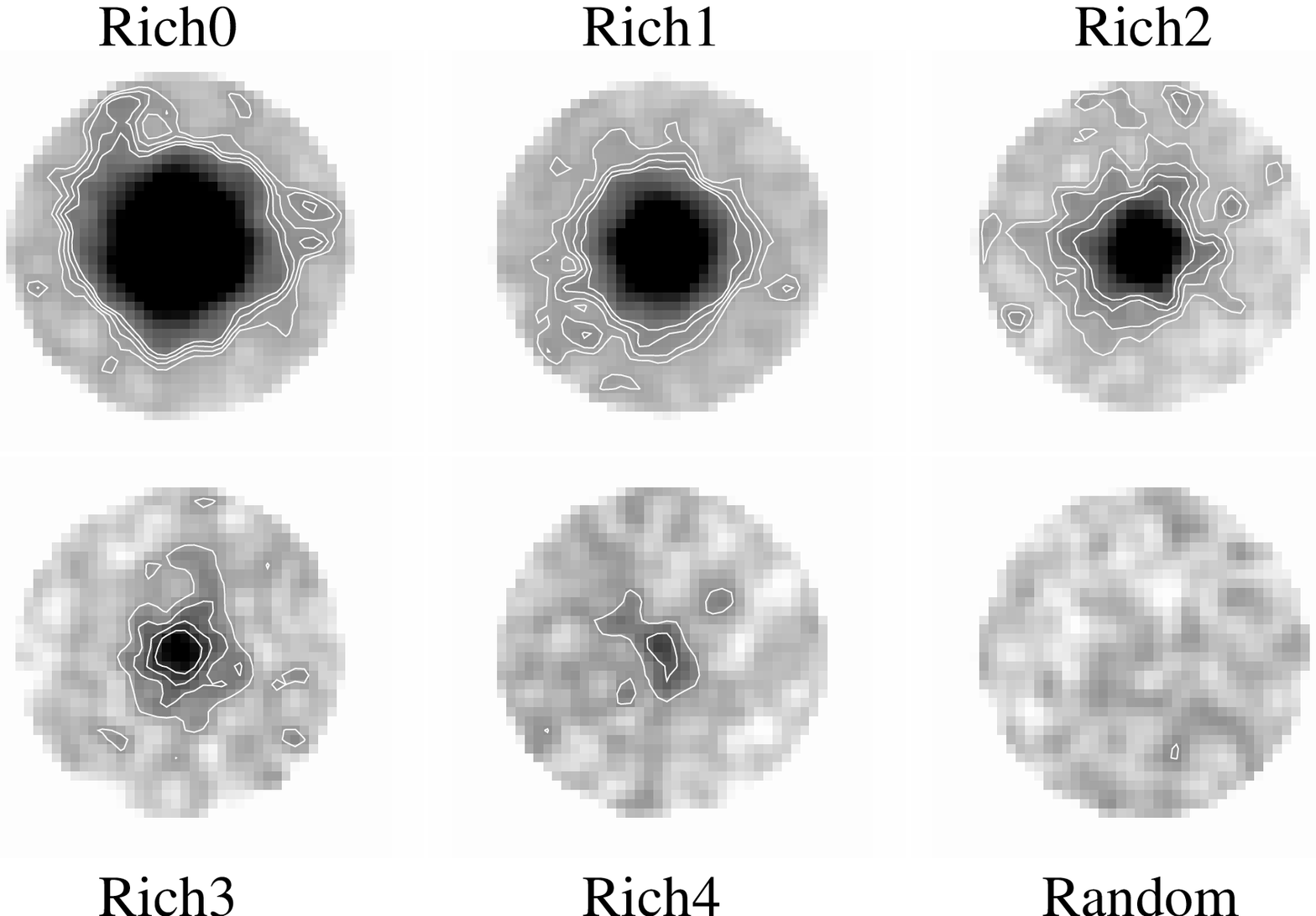}
\caption{The stacked \rosat\ images of 2MASS clusters in richness bins 0--4 and at random positions.  
The images are 2.0, 2.0, 1.5, 1.0, 1.0, and 1.0 \h Mpc in radius for richness bins 0--4 and the 
random positions.  The contours are drawn at levels of 1--4~$\sigma$ above the background.  
Excess X-ray emission is clearly detected in all richness bins, whereas the stacked image using 
random positions does not show any significant emission at the center. \label{fig:img}}
\end{figure}

\section{The Cluster Catalog and The X-ray Data}

The cluster catalog we use was selected from the 2MASS infrared survey (Skrutskie et al. 2006)
using the matched filter algorithm of Kochanek et al. (2003) applied to the 380,000 galaxies 
with K$ \leq 13.25$~mag 
(2MASS 20~mag/arcsec$^2$ circular, isophotal magnitudes) and Galactic latitude $|b|>5^\circ$.  
For the purposes of this paper we are interested only in the richness $N_{*666}$ and the
redshift $z$ of the cluster.  
The richness $N_{*666}$ is defined to be the number of 
$L>L_*$ galaxies inside the (spherical) radius $r_{*666}$ where the galaxy over density is 
$\Delta_N=200 \Omega_M^{-1} \simeq 666$ times the mean density based on the 
galaxy luminosity function of Kochanek et al. (2001).  This radius would match the
radius with a mass over density of $\Delta_M=200$ for $\Omega_M=0.3$ and a bias
factor of unity.  Clusters were included in the final catalog if they had likelihoods $\ln L \geq 10$ and
if the cluster redshift was below that at which the cluster would contain one galaxy inside
$r_{*666}$ at the magnitude limit of the survey.\footnote{The actual number of galaxies is larger
because the matched filter uses galaxies out to the smaller of a projected radius of 
$1h^{-1}$~Mpc  and $4^\circ$.  The $N_{*666}>1$ limit roughly corresponds to 3 or more galaxies
inside the filter area.  While we deliberately pushed to the completeness limits, 
almost all these clusters will be real.}
 These selection limits were determined
by examining how cluster likelihoods changed as a function of the limiting magnitude of the
input galaxy catalog.  We divided the cluster catalog into five richness bins, 
$N_{*666}\geq 10$, $10 > N_{*666} \geq 3$, $3 > N_{*666} \geq 1$ and $1 > N_{*666} \geq 0.3$,
which we will refer to as richness bins 0 (richest) to 4 (poorest) respectively.  
There are 191, 862, 1283, 1298 and 699 clusters in the bins.
We matched our cluster catalog to the Abell cluster catalog and found most matches are from 
our richness bins 0, 1, and 2.

We use the RASS to determine the average properties of 
the optically-selected clusters, since it is the only recent X-ray survey with the necessary
sky coverage.  Pointed
observations with \rosat, \chandra, or \xmm\ are not useful because they would include 
only a small fraction of the objects in our catalog and represent a biased sampling of the
catalog.   The RASS data are fairly uniform in both integration time and the average 
point spread function because the data were obtained by scanning the satellite over the
sky.  The RASS data and the exposure maps, arranged in $6.4\times6.4$ square degree fields, 
were obtained through the High Energy Astrophysics Science Archive Research Center (HEASARC\footnote{
http://heasarc.gsfc.nasa.gov/}).
We extracted the X-ray data in
fixed physical regions around the clusters of 2.0, 2.0, 1.5, 1.0, and 1.0 \h Mpc radius for 
richness bins 0--4.  These scales were chosen to be large enough to estimate the mean 
X-ray backgrounds since the background must be subtracted for our analyses of the X-ray
images and spectra.   We ignored the 17\% of the clusters for which the extraction regions
extended outside the edges of the standard RASS images.  The typical exposure time for 
each cluster is $\sim 400$~sec, and we dropped the $\sim 1\%$ of clusters with exposure
times exceeding $1500$~sec (mostly near the poles of the RASS survey) so that they would
not dominate the signal-to-noise ratio of the average.  Since we are testing the optical
catalog, dropping clusters because of the characteristics of the X-ray survey should have
no consequences for our results.  We did not distinguish between the different
observing stripes of the RASS, which have different observing times and backgrounds,
instead simply using the average exposure time and background of the combined data.
These differences were usually small (less than 10\%) and should have little effect
on our analysis.  Some examples of the optical and X-ray clusters are shown in Figure~\ref{fig:oxplot}.

We must also remove contaminating sources from each cluster field.  We removed the bright
point sources in the RASS bright source catalog (Voges et al. 1999) and replaced them with the
average of the regions surrounding them (annuli with inner and out radii of 1.5 and 2.5 times the 
extent of the point sources).  
As for the bright extended sources detected in the RASS bright source catalog, in most cases they 
are the clusters we need to stack, although we identified and excluded two clusters with 
emission from supernova remnants in our extraction region.  Finally, approximately 4\%
of the clusters had additional optical clusters in the extraction annulus.  
In these cases, 
we only analyzed the richest cluster.   
In the cases (five in total) where two clusters are of comparable richness, we excluded both of them as the optical positions tend to be poorly estimated, and 
sometimes, the optical detection algorithm only found one cluster, locating the optical centroid between the two X-ray clusters.  One such example is 
shown in Figure~\ref{fig:oxplot}d.  After applying all these filters, we were left 
with 157, 670, 1004, 1057, and 497 clusters for richness bins 0--4, respectively.

\begin{figure}
\epsscale{1}
\plotone{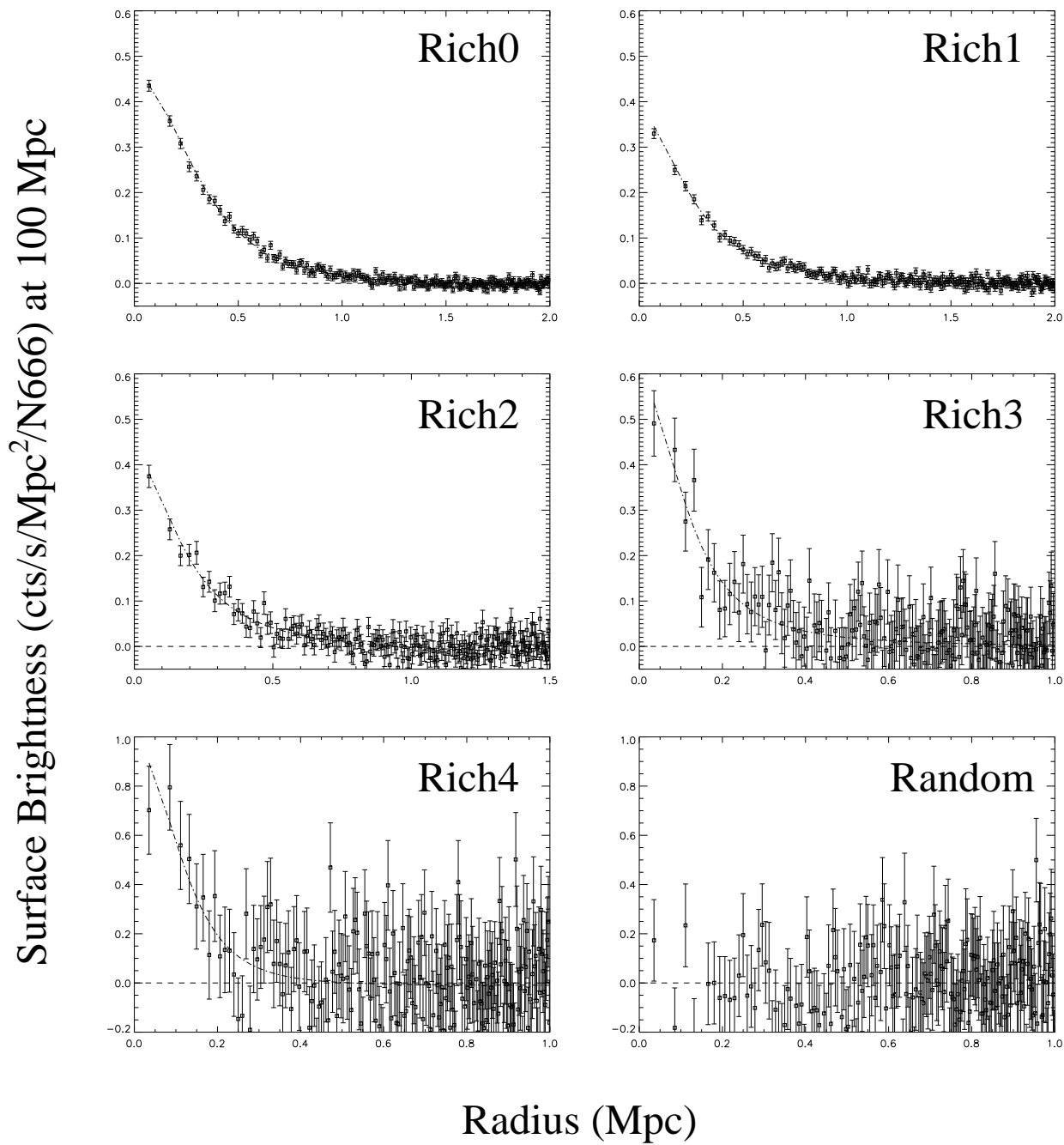}
\caption{The stacked X-ray surface brightness profiles as a function of richness and for random
 positions.  All clusters were background subtracted and normalized to a common distance of 100 Mpc 
 and by their $N_{*666}$ values.  The dashed lines indicate the level of the subtracted background and the
 dash-dotted lines are the best fit $\beta$ models for each bin. \label{fig:sb}}
\end{figure}

\begin{figure}
\epsscale{0.7}
\plotone{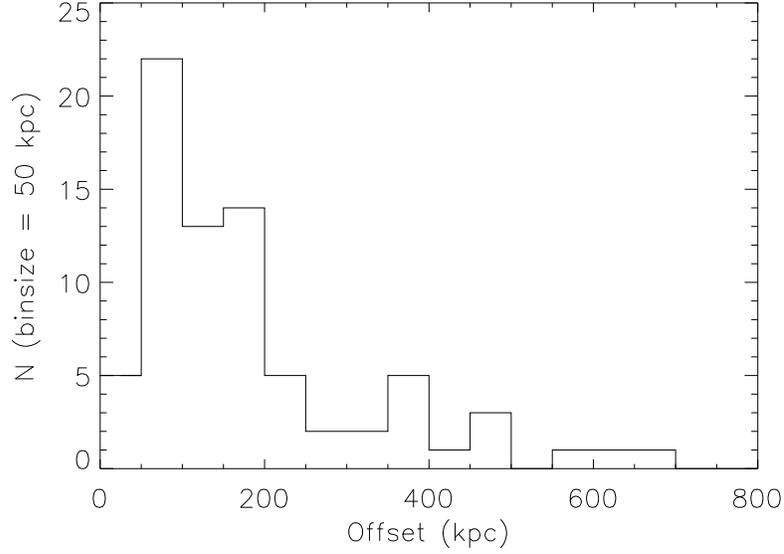}
\caption{Histogram of offsets between optical and X-ray centroids.\label{fig:offset}}
\end{figure}

\begin{figure}
\epsscale{1}
\plotone{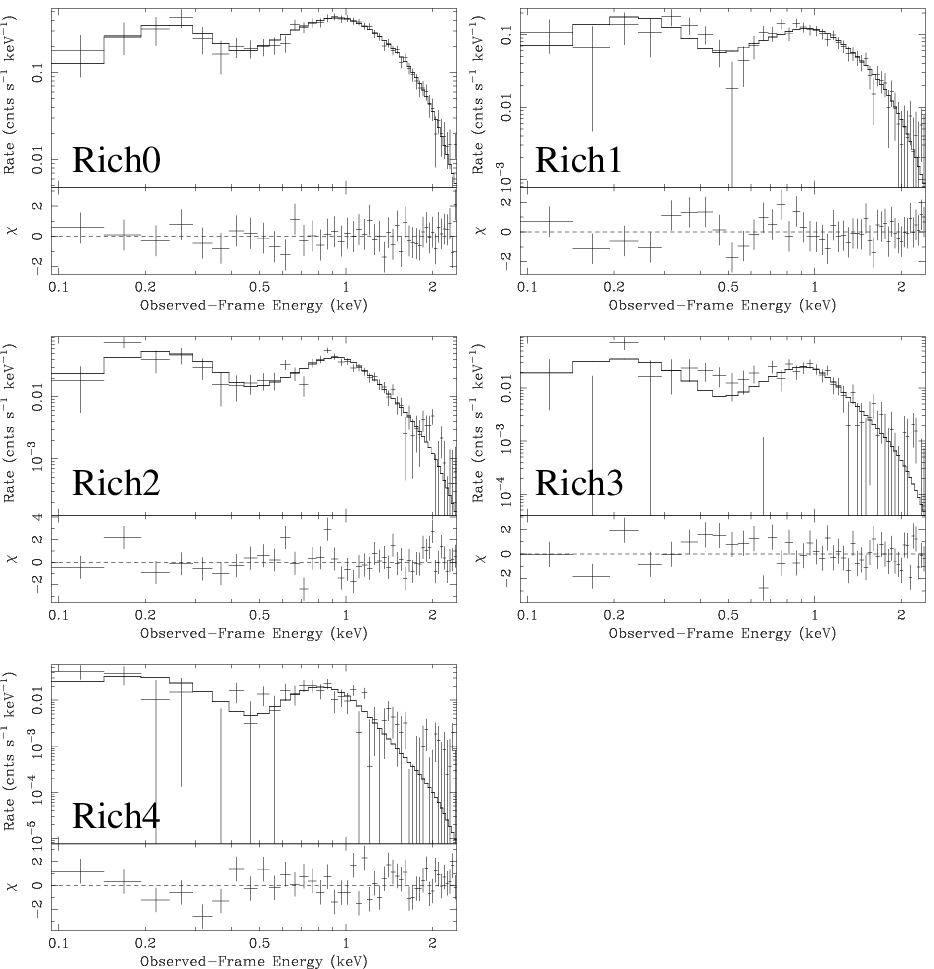}
\caption{The stacked \rosat\ spectra of clusters in richness bins 0--4 and their best fit thermal plasma models.\label{fig:spec}}
\end{figure}

\subsection{Surface Brightness Profiles \label{sec:sbright}}

We produce the stacked images by
rescaling the data for each cluster to a common distance of 100 Mpc.  In addition to adjusting
the photon positions, we weight each photon by the square of the ratio between the cluster
distance and 100~Mpc.  We binned the X-ray images with pixel sizes of 100, 100, 75, 50, and
50 \h kpc for richness bins 0--4 and then smoothed them with a Gaussian with a dispersion of
one pixel.   We clearly detect X-ray emission from the clusters in all five richness bins,
as shown in Figure~\ref{fig:img}.
The significance contours in Figure~\ref{fig:img} apply to the individual pixels, so the sources as a whole are detected at
much higher confidence levels than indicated by the contours.
For example, although the peak of the emission from the poorest richness bin is only $2\sigma$ above the background,
the emission as a whole is detected at about $7\sigma$.  We also used bootstrap re-sampling to
test the significance of the X-ray emission of the poorest richness bin.  First, we tested
whether the emission was dominated by a small subset of the objects by randomly drawing identically
sized samples from the original sample with replacement and then analyzing the simulated sample.
In each of the 100 trials there was excess X-ray emission detected at over 3$\sigma$, and in 94 out of 100
trials the cluster was detected at over 5$\sigma$. If, on the other hand, we stack data from an equal number of
random positions with $|b|>5^\circ$ and excluding our standard regions near the 2MASS cluster
positions, only 8 out of 100 trials have 3$\sigma$ detections and no trials are detected at
over 4$\sigma$.  One such image is shown in Figure~\ref{fig:img}.  Based on these simulations,
we estimate that we detect the richness 4 clusters at greater than the 99\% confidence level.
In the outer, background regions of the images, the different richness bins have consistent
background flux levels and when azimuthally smoothed they show the flat profile expected for
convergence.

The background-subtracted surface brightness profiles for each richness bin are shown in Figure~\ref{fig:sb},
where we have used radial bins of constant area, normalized the surface brightness profile by the optical
richness $N_{*666}$ and scaled the results to a luminosity distance of 100 Mpc.  We subtracted the
background for each cluster before we rescaled the data. Like the stacked images, the surface brightness
profiles clearly show excess emission above the background for all richness bins, while the profile from
stacking random positions does not.  We fit the profiles with a standard $\beta$ model,
$S(R) = S_0(1+R^2/R_c^2)^{-3\beta+1/2}$ obtaining the results presented in Table~\ref{tab:betafit}.
We obtain similar results whether we fit the background-subtracted profiles directly or include
an additional parameter to fit any residual background level.  When we include an additional
parameter for the residual background, its value is found to be consistent with zero.
Our estimates of the slope $\beta$ are consistent with typical values for clusters (e.g., Xu et al. 2001; Sanderson et al. 2003),
even consistent with the trend of decreasing $\beta$ with decreasing cluster temperature (e.g., Horner et al. 1999; Sanderson et al. 2003)
for richnesses 0--3.  The $\beta$ values obtained for the two poorest richness bins have large
uncertainties.  We do, however, find larger core sizes than studies of individual clusters
(e.g., Neumann \& Arnaud 1999; Xu et al. 2001; Sanderson et al. 2003; Osmond \& Ponman 2004),
ranging from roughly twice as large for richness bin 0 to more than 5 times as large for richness
bin 4,  although the estimates for the poorest richness bins are very uncertain.
This is not due to the RASS PSF, which is equivalent to a source size of $\sim 20$ kpc at a
typical cluster redshift of $z \sim 0.04$.   Part of the problem must be a smearing of
the X-ray cores by errors in the estimates of the optical centroids.  We matched our stacked clusters with the 
\rosat\ extended source catalog and plotted the histogram of offsets between optical and X-ray centroids in 
Figure~\ref{fig:offset}.  The median offset of 0.14 Mpc is consistent with the large core sizes obtained 
for our stacked images in richness bins 0 and 1 compared with those obtained from individual clusters.
The distribution of offsets is very similar to other studies (e.g., Adami et al.\ 1998; Lopes et al.\ 2006) when
comparing optically-selected clusters and ROSAT clusters.
Some of the large offsets are caused by merging clusters where the X-ray image has two cores and the optical 
centroid is set in 
between of the X-ray cores (e.g., Figure~\ref{fig:oxplot}b), and in some other cases the large offset could be due to various other 
problems (e.g., Figure~\ref{fig:oxplot}c).  We also show one case where the optical centroid matches excellently 
with the X-ray centroid in Figure~\ref{fig:oxplot}a.
Although we obtain larger cores in our stacked images, the average X-ray properties of the clusters should be little affected.

\subsection{Cluster Luminosities, Temperatures and Entropies}

In order to estimate the luminosities and temperatures we need to model the X-ray spectra of the
clusters.  We started by eliminating the 13\% of clusters
with Galactic absorption column densities higher than $10^{21}$\cmsq.
We analyzed the spectra using \verb+XSPEC+ (Arnaud 1996) using the \verb+rmf+ file for
the PSPCC detector available from HEASARC.  Although the RASS contains some PSPCB observations, they are a small
minority of the observations and the \verb+rmf+ files for the two detectors are very similar.
Because the RASS was carried out by scanning the sky, there is one
additional complication arising from the corrections for vignetting
and the off-axis telescope response.  We can use two possible
calibration files for our analyses.  The first approach is to
use the standard \verb+rsp+ file provided in the calibration package
(which is equivalent to the on-axis \verb+arf++\verb+rmf+ files) combined with
the vignetting-corrected exposure map.  This approach will
bias the spectral fits because the true vignetting correction
is energy-dependent.  The second approach is to use the
off-axis \verb+arf+ file. This will lead to unbiased spectral fits, but
the RASS images have already corrected the exposure times for
the effects of vignetting, so we will be double-counting this
correction when we use the off-axis \verb+arf+ file.  Fortunately, this
simply leads to a normalization offset in the 0.1--2.4~keV flux
of a constant factor of $1.44$ that we can determine from the
flux differences between the two approaches.  We adopt this latter
approach with the correction to the fluxes.
We generated the off-axis \verb+arf+ files using the \verb+pcarf+ software tool from the \verb+XSELECT+ and \verb+FTOOLS+ packages.
We extracted the spectra of the stacked clusters out to radii of 1.5, 1.5, 0.9, 0.65, and 0.4 Mpc for
richness bins 0--4, respectively.  For each cluster we subtracted the background and normalized the
data to the mean redshift of the stacked clusters ($z=0.0816$, 0.0678, 0.0539, 0.0387, and 0.0273 for
richness bins 0--4) and then averaged the \verb+arf+ files based on the weighted number of photons
in each cluster.  Averaging of \verb+arf+ files has little effect on the results because they are
all fairly similar due to the nature of the RASS observations.

We fit the spectra with a thermal plasma model (Raymond \& Smith 1977) modified by the mean Galactic
absorption. We were unable to determine the metallicity from the data, so we assumed a metallicity
of $1/3$ solar (e.g., Baumgartner et al. 2005).   The redshift was set to the mean redshift of the clusters
in each richness bin.  We did not worry about the small spread in the true K-corrections for
the spectra across each bin because the low redshift of the sample (see Table 1) makes this a
small effect.  The spectra and their best fit models are shown in Figure~\ref{fig:spec} and the
fitting results are presented in Table~\ref{tab:spec}.  We obtained acceptable fits for all
richness bins, indicating that the emission is consistent with that from hot gas for all richness
bins.   As expected, the temperature decreases from rich to poor clusters, and the
soft X-ray sensitivity of \rosat\ (0.1--2.4~keV) allows us to measure the temperature of the poor clusters
more accurately than for rich clusters.  Given the temperatures, we can also estimate the
bolometric X-ray luminosity.  Table~\ref{tab:spec} presents the estimate for the luminosity
within $R_{*666}$ found by using the $\beta$ model fits from \S2.1 to correct from the spectral
extraction aperture to $R_{*666}$.   These corrections were quite small, amounting to
multiplicative factors of only 1.06, 0.92, 1.01, 0.99, and 1.03 for richness bins
0--4, respectively.  In addition to the error estimates provided by the spectral fits, we
also estimated the uncertainties by bootstrap re-sampling the data.  For each richness
bin we generated 500 random samples by drawing the same number of clusters from
those in each richness bin with replacement.  These bootstrap estimates of the uncertainties should represent
systematic uncertainties better than the purely Poisson uncertainties of the standard fits.
In general, the bootstrap uncertainties are slightly larger than the standard errors found from the
spectral fits, but usually not by large margins.

While the best fit Galactic \nh\ values are consistent with our \nh\ cut of $10^{21}$\cmsq, there
is a worrisome correlation of \nh\ with richness.  The estimate of the Galactic absorption is
dominated by the lowest energy bins.
  We ran
several fits with fixed \nh\ values in the range spanned by the results of Table~\ref{tab:spec},
finding that the overall fits become worse for the richness bins best fit by a different
value for \nh, but that the luminosity and temperature estimates change little compared to the statistical errors of the original estimates.
We found that we could not analyze the high
Galactic absorption clusters at all because of strong degeneracies between the estimates
of the temperature and the absorption.   Potentially, the origin of the problem could be
the presence of an additional soft emission component in the low richness clusters, but
we will not investigate this possibility here because it has no effect on our general results and
is an area of considerable controversy (e.g., Bregman \& Lloyd-Davies 2006).

We also estimated the mean gas entropy, $S=T/n_e^{2/3}$, as a direct test of the thermodynamic state 
of the clusters and the effects of heat transport or energy injection 
(e.g., Ponman, Cannon, \& Navarro 1999;
Lloyd-Davies, Ponman, \& Cannon 2000; Ponman et al. 2003). We estimated 
the electron number density $n_e$ from the surface brightness profiles derived in \S\ref{sec:sbright}
and the temperature from the spectral fits.  We assumed a constant temperature with radius.
As we discussed in \S\ref{sec:sbright}, we systematically obtain larger core radii than analyses of individual
clusters, and an underestimate of the central surface brightness will lead to an overestimate
of the central entropy.  Therefore, we focus on the entropy on scales of $0.5 r_{*666}$ where we
should be insensitive to this problem. 

We further divided the each of the first four original bins into two narrower bins.  We followed the same
procedure described in this section to obtain their temperatures and the bolometric X-ray luminosities 
(Table~\ref{tab:spec}).

\subsection{Cluster Masses \label{sec:cmass}}

Finally, we estimated the mean cluster masses inside $r_{*666}$ assuming hydrostatic equilibrium,
\begin{equation}
M_{grav}(<r) = -\frac{kT(r)r}{G \mu m_p}\left(\frac{d\ln{\rho(r)}}{d\ln{r}}+\frac{d\ln{T(r)}}{d\ln{r}}\right),
\end{equation}
using our $\beta$-models from \S\ref{sec:sbright}, $\mu = 0.61$ (Arnaud 2005a), and assuming an isothermal temperature profile.
This is not a trivial exercise given the stacked data.  First, we cannot accurately measure the mass profile
in the inner regions of the clusters where we can accurately measure the temperature because of the smearing
created by the position uncertainties discussed in \S2.1.  Second, we cannot accurately measure the temperature
in the outer regions, or equivalently, the slope of the temperature profile, because we have insufficient
counts in the outer regions.  
We expect the first issue is less severe since the mass obtained by this method mainly depends on the temperature 
and its gradient at a sufficiently large radius, which is less affected by the profile of the inner slope.
The isothermal assumption will possibly lead to an over estimate of mass by $\sim$15\% (Sanderson et al.\ 2003).
These uncertainties are smaller than those introduced from cluster temperature measurements for most cases.
We estimated $r_{200}$ and $M_{200}$ from the mass profiles by calculating 
the mass and radius where the mass over density is 200 times the critical density. 
We also performed the same analysis to the narrower richness bins.  Again, we expect the dominant uncertainties
for masses are from the cluster temperatures.

\begin{figure}
\epsscale{0.7}
\plotone{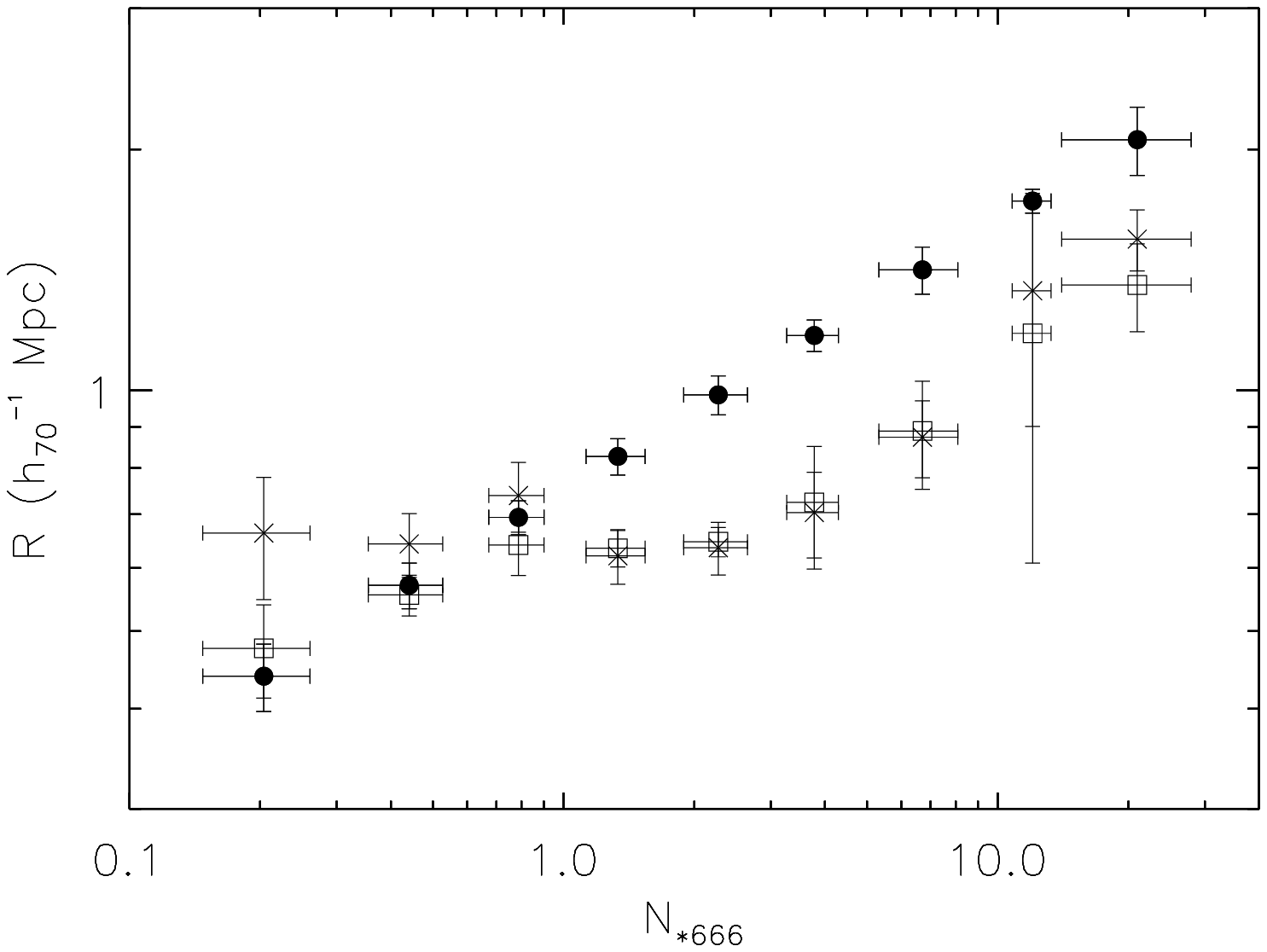}
\caption{The relationship between cluster scaling radius and optical richness $N_{*666}$.  The filled circles are 
  $r_{*666}$ obtained from optical algorithm, the squares are $r_{180}$ converted from mean temperature 
  (Evrard et al. 1996), and the X symbols are $r_{200}$ calculated from our mass profile for stacked
  clusters.  The radius $r_{*666}$ is algebraically related to $N_{*666}$ because of the fixed model 
  for the spatial distribution of the galaxies (see Kochanek et al. 2003). \label{fig:r}}
\end{figure}

There are offsets between the $M_{*666}$ and $M_{200}$ mass estimates that simply reflect the differences between
$r_{*666}$ and $r_{200}$.  The two radii are not simply proportional to each other, 
as illustrated in Fig.~\ref{fig:r}.
The difficulty probably arises because $r_{*666}$ is computed from the fitted richness and a fixed three
dimensional galaxy distribution because it is not possible to determine the galaxy distribution for the
individual clusters with any accuracy (see Kochanek et al. 2003). To investigate this further we need to
examine the optical properties of the stacked clusters, which will be the subject of a later study
(Morgan et al. 2007 in preparation).  Our
estimate of $r_{200}$ from the stacked X-ray images is consistent with scaling relations from the
measured temperatures, as illustrated by the values of $r_{180}$ derived from the temperature and
the temperature-radius scaling relation from Evrard et al. (1996). Note, however, that the shifts in the masses
due to the different radii are too small to affect the overall structure of the mass-temperature-richness correlations.

\section{Averaging Cluster Properties at Fixed Optical Richness}

Before discussing the results from measuring the mean luminosity, temperature or mass as a function
of optical richness, it is necessary to discuss the important distinction between averaging at fixed
richness and averaging at fixed mass (see Berlind \& Weinberg 2002).  For simplicity, we will assume a 
simple model for the halo mass function,
\begin{equation}
   n_M(M) = { dn \over dM } \propto M^{-x} \exp\left(-(M/M_1)^y \right)
\end{equation}
with $x \simeq 1.88$ and $\log (M_1/M_\odot) \simeq 14.6$ and $y \simeq 0.73$ closely
reproducing a concordance cosmology mass function at $z=0$.
 Since only higher mass halos form groups and clusters of massive galaxies (our cluster finder
is not designed to find galaxies with halos of lower luminosity satellites), we must truncate the
halo mass function $n_M(M)$ at some point for it to model the cluster mass function $n_c(M)$.  We model this
by using $n_c(M)=n_M(M)$ for $M>M_0$ and then truncating it at lower masses
\begin{equation}
   n_c(M) = n_M(M_0) \exp\left( - { \log^2 \left(M/M_0\right) \over 2 \Delta^2 }  \right) \quad\hbox{for}\quad M<M_0
\end{equation}
over a logarithmic mass range $\Delta$.  On average, a cluster of mass $M$ contains 
\begin{equation}
     \langle n \rangle = N_1 \left( { M \over M_1 } \right)^\gamma { \Gamma[1+\alpha,L(z)/L_*] \over \Gamma[1+\alpha,1] }
              = N_1 \left( { M \over M_1 } \right)^\gamma c(z)
\end{equation}
{\it observable} galaxies, where $N_1$ is the number of $L>L_*$ galaxies for a cluster of mass $M_1$
and the ratio of Gamma functions $c(z)$
corrects from the number of $L>L_*$ galaxies to the number of galaxies down to the luminosity $L(z)$ 
corresponding to the flux limit of the galaxy survey at the cluster redshift $z$.  The galaxy luminosity function 
is modeled as a Schechter function with slope $\alpha=-1.09$ and break luminosity $M_{K*} = -24.16$~mag assumed by 
the cluster finder based on the Kochanek et al. (2003) K-band luminosity functions and with a limiting magnitude of 
$K_{lim}=13.25$~mag.  A particular cluster, however, will contain $n$ galaxies drawn from the
conditional probability function for a cluster of mass $M$ having $n$ galaxies,
\begin{equation}
      P(n|M) = { \langle n \rangle^n \over n! } \exp(-\langle n \rangle)
\end{equation}
which we assume to be the Poisson distribution.  In general, the cluster catalogs are roughly 
complete to the limit of $n \geq 3$ inside the radius of the filter (see White \& Kochanek 2002; 
Kochanek et al. 2003), and this turns out to be the best fit estimate of the selection limit
when we apply the Poisson model to the data. The 
richness estimate assigned to the cluster is not the observed number of galaxies $n$, but
the estimated number of $L>L_*$ galaxies, $N_*=n/c(z)$.  For our
analytic models, we will further assume that there are perfect correlations $T(M)$ and $L_X(M)$
between the X-ray properties and the mass.  The combination of these assumptions is essentially
a simple example of a halo occupation distribution (HOD) model (e.g. Yang et al. 2005, Zheng et al. 2005) for the clusters.

When we stack the X-ray data for cluster of fixed observed richness, $N_*$, we compute mean values
of the form, given here for the mean mass,
\begin{equation}
   \langle M(N_*,z)\rangle = { \int_0^\infty dM M n_c(M) P(N_* c(z) |M)  \over
                            \int_0^\infty dM  n_c(M)  P(N_* c(z) |M) }.
\end{equation}
If we set $\gamma=1$, so that the number of galaxies is proportional to the mass, and use a sharp
cutoff $\Delta=0$ in the cluster mass function, then the integrals can be done analytically to 
find that
\begin{equation}
  \langle  M (N_*,z)  \rangle = 
   { M_1 \over 1+ N_1 c(z) } { \Gamma[2-x+N_*c(z), { M_0 \over M_1} (1+N_1 c(z))] \over 
          \Gamma[1-x+N_*c(z), { M_0 \over M_1} (1+N_1 c(z))] }
\end{equation}
where $\Gamma[a,b]$ is an incomplete Gamma function.  If we had averaged at fixed mass, we
would simply find that the mean mass is $M(N_*)=M_1(N_*/N_1)$, and it is interesting to 
examine how the average at fixed richness differs from that at fixed mass.  If we 
consider rich clusters with many galaxies ($M \gg M_0$, $N_* c(z) \gg 1$) then we find that 
\begin{equation}
 { \langle  M (N_*,z)  \rangle \over M(N_*) } = \left[1 + { 1 - x \over N_* c(z)}\right]
                         \left[ 1 + {1 \over N_1 c(z) } \right]^{-1}.
\end{equation}  
The Poisson fluctuations bias the average mass at fixed richness to
be low, with a fractional amplitude of order the number of detectable galaxies in the 
cluster.  This is essentially a Malmquist bias due to the steep mass function -- 
more low mass clusters Poisson fluctuate up to the fixed richness than high mass 
clusters fluctuate down to it.  The more interesting limit is that of very low
richness clusters.  Very low richness clusters will always be close to the cluster 
detection limit, so let us set $N_* c(z)=3$ and take the limit of very low richness 
($N_* \rightarrow 0$) to find that 
\begin{equation}
   \langle  M (N_*,z)  \rangle \rightarrow M_0.
\end{equation}
The only way to get an apparently very low richness cluster is as a Poisson fluctuation of 
a cluster with $M \simeq M_0$, so the mean mass converges to $M_0$.  Thus, compared to
averaging properties of clusters at fixed mass, clusters averaged at fixed luminosity
should be biased lower in mass for rich clusters and higher in mass for poor clusters
even before considering any selection effects.  As we shall see, this simple model
explains all the results that follow.

\begin{figure}
\epsscale{1}
\plotone{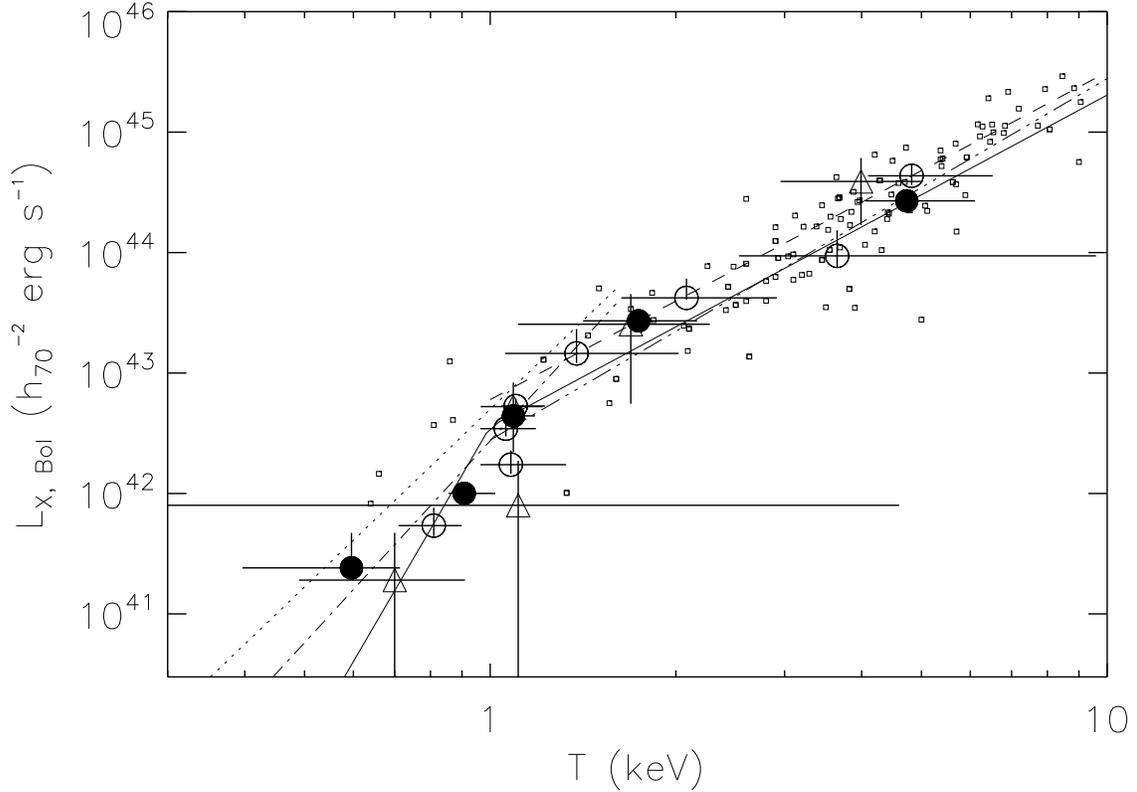}
\caption{The relationship of the bolometric X-ray luminosity to the temperature.
  The open and filled circles are the results from our stacking the clusters as a function
  of richness, and the solid line is the best fit Poisson model (see \S3) to these results.   
  The filled circles are 
  for the five standard richness bins, while the open circles are the results found for 
  the narrower bins obtained by dividing the first four richness bins in half. 
  The open triangles are the results found for the middle redshift bin (see Figs.~\ref{fig:lz}, \ref{fig:tz}) when we divide 
  the clusters in redshift sub-bins.
  The error bars on the circles are the bootstrap estimates of the errors.
  The squares show the results for individual 2MASS clusters drawn from the literature.
  The dashed, dotted, dash-dotted, and dash-dot-dot-dotted lines are from 
  the $L_X$--$T_X$ relations from Wu et al. (1999), Helsdon \& Ponman (2000), and 
  Xue \& Wu (2000), and Rosati et al. (2002).  
  \label{fig:lt}}
\end{figure}

\begin{figure}
\epsscale{1}
\plotone{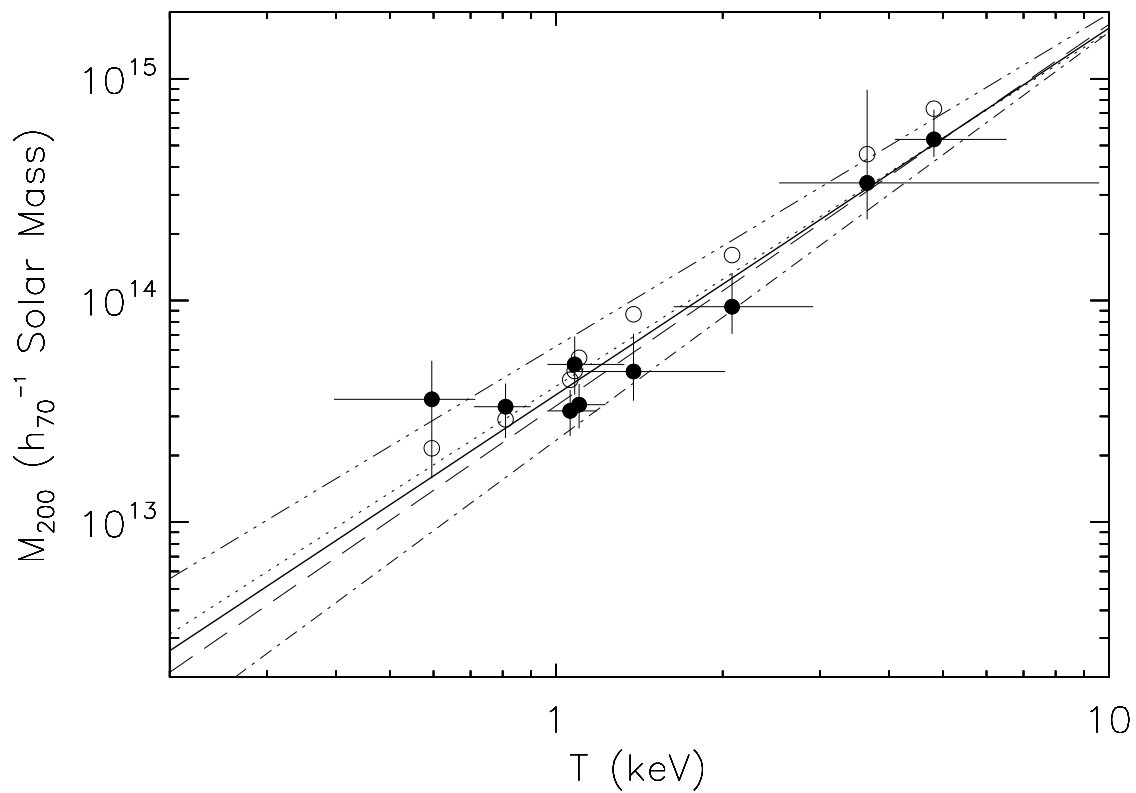}
\caption{The relationship between mass $M$ and temperature $T$, using the subdivided bins for the first four
  richness bins.
   The filled circles are the $M_{200}$ values for the stacked clusters, and the solid line
   is our best fit Poisson model (see \S3).  The open circles are the $M_{*666}$ values, and the error-bars of the open circles are similar to those for the filled circles.
   The dashed, dotted, dash-dotted, and dash-dot-dot-dotted lines are the relations from Arnaud et al. (2005b), Xu et al. (2001), Sanderson et al. (2003), and Neumann \& Arnaud (1999).
  \label{fig:mt}}
\end{figure}

\begin{figure}
\epsscale{0.7}
\plotone{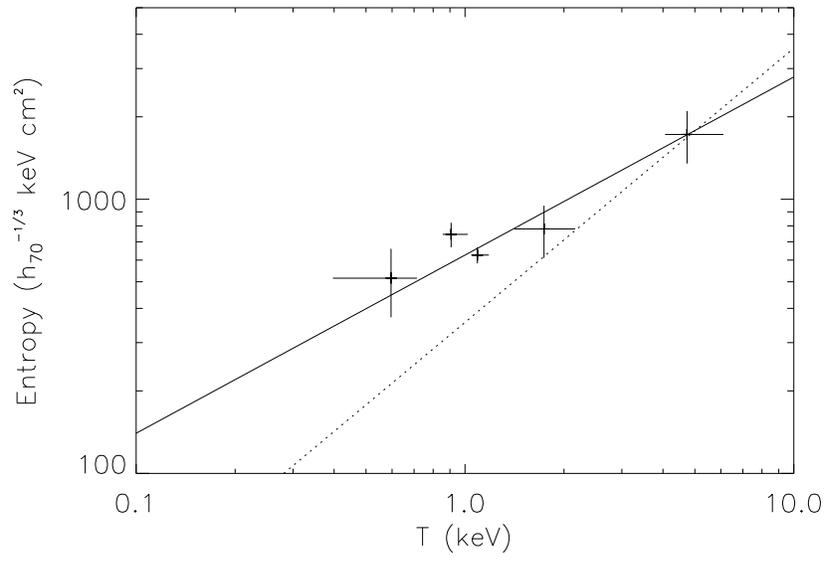}
\caption{The relationship between temperature and entropy at 0.5 $r_{*666}$.  The slope of the solid line
   is the $S \propto T^{0.65}$ relation from Ponman et al. (2003) and the dotted
   line is the $S \propto T$ relation expected from simple scaling relation.
  \label{fig:en}}
\end{figure}

\begin{figure}
\epsscale{1}
\plotone{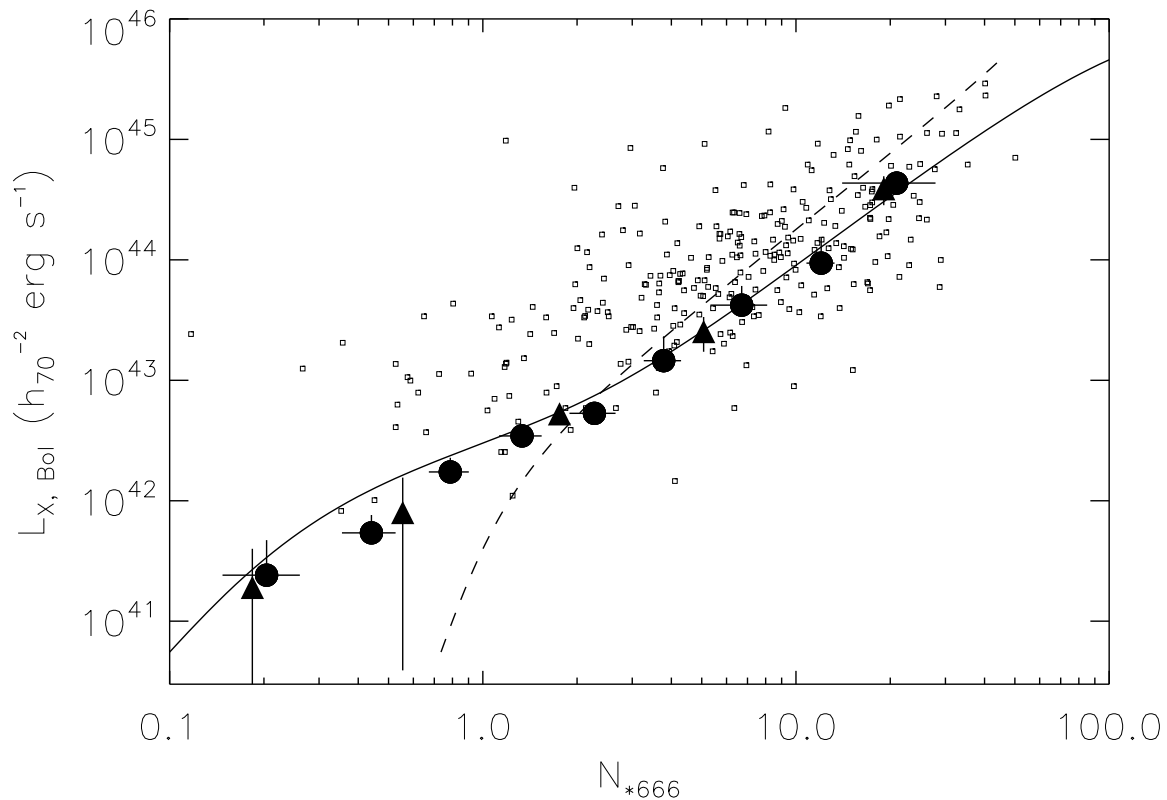}
\caption{The relationship of the bolometric X-ray luminosity to the optical richness $N_{*666}$.
  The filled circles with error-bars are our present results, using the subdivided bins for the first four
  richness bins.  The filled triangles are data from the middle redshift bin (see Figs.~\ref{fig:lz}, \ref{fig:tz}).
  The solid line shows the best fit Poisson model to the data, and the dashed line shows the true
  relationship between $L_X$ and $N_{*666}$ we would observe in the absence of Poisson fluctuations.
  The squares are individual 2MASS clusters drawn from the literature.
  \label{fig:ln666}}

\end{figure}

\begin{deluxetable}{ccccc}
\tabletypesize{\scriptsize}
\tablecolumns{5}
\tablewidth{0pt}
\tablecaption{Average Cluster Surface Brightness Profiles \label{tab:betafit}}
\tablehead{
\colhead{} &
\colhead{} &
\colhead{} &
\colhead{$R_c$} &
\colhead{} 
\\
\colhead{Richness} &
\colhead{$N_{*666}$ range} &
\colhead{redshift range} &
\colhead{(\h Mpc)} &
\colhead{$\beta$} 
}
\startdata
0 & 10--50 & 0.018--0.12 & $0.43_{-0.03}^{+0.02}$ & $0.70_{-0.05}^{+0.05}$ \\
1 & 3--10  & 0.004--0.12 & $0.27_{-0.04}^{+0.04}$ & $0.51_{-0.05}^{+0.07}$ \\
2 & 1--3   & 0.005--0.10 & $0.19^{+0.05}_{-0.07}$ & $0.48_{-0.10}^{+0.09}$ \\
3 & 0.3--1 & 0.003--0.076 & $0.14^{+0.08}_{-0.07}$ & $0.6^{+0.3}_{-0.2}$ \\
4 & 0.1--0.3 & 0.003--0.049 & $0.20_{-0.08}^{+0.09}$ & $0.9_{-0.3}^{+0.5}$ \\
\enddata

\end{deluxetable}

\begin{deluxetable}{cccccccc}
\tabletypesize{\scriptsize}
\tablecolumns{8}
\tablewidth{0pt}
\tablecaption{Spectral Fitting Results \label{tab:spec}}
\tablehead{
\colhead{} &
\colhead{} &
\colhead{$R_{*666}$} &
\colhead{} &
\colhead{Galactic \nh\ } &
\colhead{$T$} &
\colhead{$L_X$ (Bol)} &
\colhead{} 
\\
\colhead{Richness} &
\colhead{$N_{*666}$} &
\colhead{(\h Mpc)} &
\colhead{$z$} &
\colhead{($10^{20}$~\cmsq)} &
\colhead{(keV)} &
\colhead{($h_{70}^{-2}\lumin$)} &
\colhead{$\chi^{2}_{\nu}(dof)$} 
}

\startdata
\cutinhead{Broad Bins}
0 & 16.57 & 1.89 & 0.0816 & $2.8^{+0.3}_{-0.3}$ & $4.7^{+1.4}_{-0.7}$    & $2.7^{+0.3}_{-0.4} \times 10^{44}$ & 0.47(44) \\
1 & 5.27  & 1.29 & 0.0678 & $1.8^{+0.3}_{-0.3}$ & $1.7^{+0.5}_{-0.3}$    & $2.7^{+0.5}_{-0.2} \times 10^{43}$ & 0.66(44) \\
2 & 1.80  & 0.91 & 0.0539 & $1.4^{+0.3}_{-0.2}$ & $1.09^{+0.09}_{-0.05}$ & $4.4^{+0.5}_{-0.5} \times 10^{42}$ & 1.24(44) \\
3 & 0.60  & 0.63 & 0.0387 & $0.7^{+0.4}_{-0.2}$ & $0.91^{+0.10}_{-0.05}$ & $1.0^{+0.2}_{-0.1} \times 10^{42}$ & 1.14(44) \\
4 & 0.20  & 0.44 & 0.0273 & $<0.4$              & $0.60^{+0.11}_{-0.20}$ & $2.4^{+2.2}_{-0.1} \times 10^{41}$ & 1.16(44) \\
\cutinhead{Narrow Bins}
0 & 20.97 & 2.06 & 0.0816 & $3.3^{+0.5}_{-0.4}$ & $4.8^{+1.7}_{-0.7}$    & $4.3^{+1.0}_{-0.7} \times 10^{44}$ & 0.63(44) \\
0 & 12.03 & 1.72 & 0.0805 & $1.5^{+0.6}_{-0.5}$ & $3.7^{+5.9}_{-1.1}$    & $9.4^{+6.3}_{-2.1} \times 10^{43}$ & 0.47(44) \\
1 &  6.71 & 1.41 & 0.0704 & $1.7^{+0.4}_{-0.3}$ & $2.1^{+0.8}_{-0.4}$    & $4.2^{+1.9}_{-0.1} \times 10^{43}$ & 0.59(44) \\
1 &  3.78 & 1.17 & 0.0651 & $1.7^{+0.8}_{-0.5}$ & $1.4^{+0.6}_{-0.3}$    & $1.5^{+0.8}_{-0.3} \times 10^{43}$ & 0.62(44) \\
2 &  2.27 & 0.99 & 0.0558 & $1.1^{+0.3}_{-0.2}$ & $1.10^{+0.13}_{-0.06}$ & $5.3^{+0.8}_{-0.7} \times 10^{42}$ & 1.29(44) \\
2 &  1.33 & 0.83 & 0.0515 & $1.7^{+0.6}_{-0.4}$ & $1.06^{+0.13}_{-0.10}$ & $3.5^{+0.6}_{-0.5} \times 10^{42}$ & 1.46(44) \\
3 &  0.79 & 0.69 & 0.0444 & $0.8^{+0.4}_{-0.3}$ & $1.08^{+0.25}_{-0.12}$ & $1.7^{+0.6}_{-0.3} \times 10^{42}$ & 0.91(44) \\
3 &  0.44 & 0.57 & 0.0354 & $1.0^{+1.6}_{-0.6}$ & $0.81^{+0.09}_{-0.10}$ & $5.4^{+2.2}_{-1.0} \times 10^{41}$ & 1.26(44) \\
4 & 0.20  & 0.44 & 0.0273 & $<0.4$              & $0.60^{+0.11}_{-0.20}$ & $2.4^{+2.2}_{-0.1} \times 10^{41}$ & 1.16(44) \\
\enddata

\tablecomments{The $N_{*666}$, $R_{*666}$, and $z$ values listed here are the average values of each bin.  The uncertainties of $L_X$ and $T$ are the bootstrap estimates.}
\end{deluxetable}

\begin{figure}
\epsscale{1}
\plotone{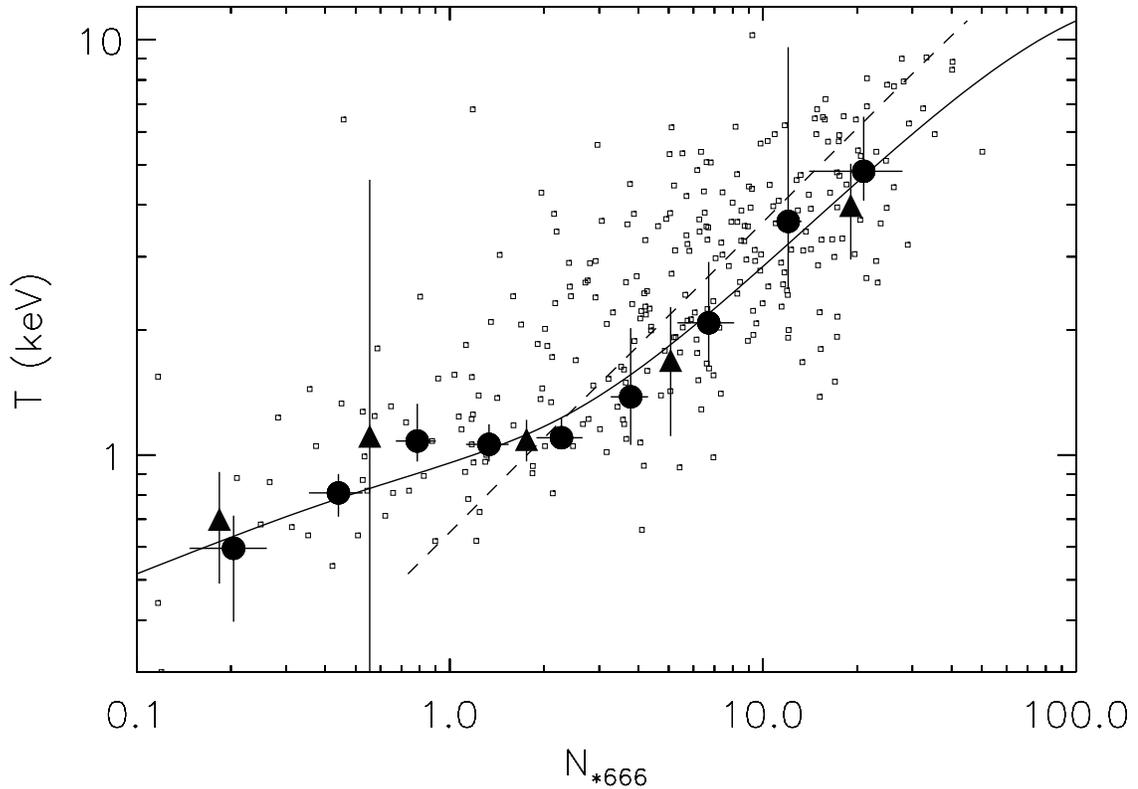}
\caption{The relationship between X-ray temperature and optical richness $N_{*666}$.
   The filled circles with error-bars are our present results, using the subdivided bins for the
   first four richness bins.  The filled triangles are for the clusters in the middle redshift bin (see Figs.~\ref{fig:lz}, \ref{fig:tz}).
  The solid line shows the best fit Poisson model to the data, and the dashed line shows the true
  relationship between $T$ and $N_{*666}$ we would observe in the absence of Poisson fluctuations.
  The squares are individual 2MASS clusters drawn from the literature.
  \label{fig:tn666}}
\end{figure}

\begin{figure}
\epsscale{1}
\plotone{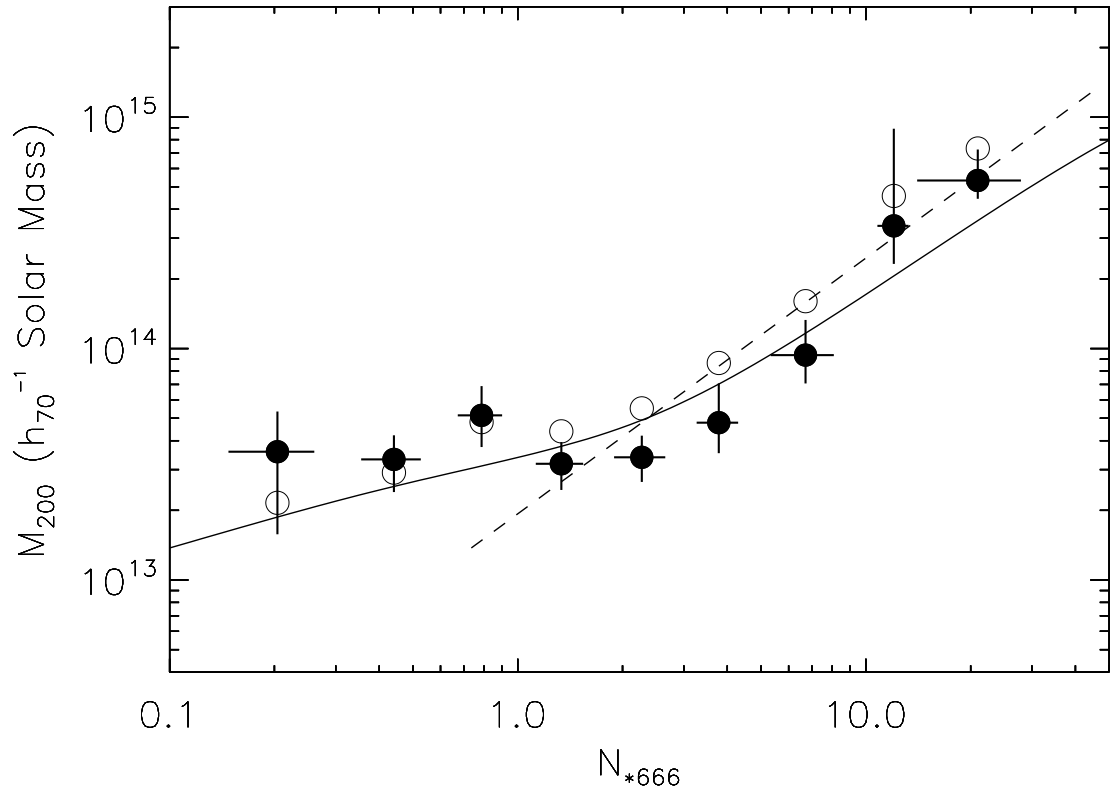}
\caption{The relationship between mass $M_{200}$ and optical richness $N_{*666}$.  The filled and open
  circles are the $M_{200}$ and $M_{666}$ values, and the error-bars for the open circles are similar to
  those for the close circles.
  The solid line shows the best fit Poisson model to the data for $M_{200}$, and the dashed line shows the true
  relationship between $M_{200}$ and $N_{*666}$ we would observe in the absence of Poisson fluctuations.
  \label{fig:mn666}}
\end{figure}

\section{Sample Correlations}

We summarize the general correlations in Figs.~\ref{fig:lt}-\ref{fig:mn666}.  We start
with the correlations familiar from studies of X-ray clusters.  Figure~\ref{fig:lt}
shows the correlation between X-ray luminosity and temperature, Figure~\ref{fig:mt}
show the relation between cluster mass and temperature, and Figure~\ref{fig:en} shows
the relation between entropy and temperature.  In each case we show the results
for the stacked clusters, the results of our global fit to the cluster properties
(the ``Poisson model'', except Fig.~\ref{fig:en})
and several similar scaling relations from the literature.  In some cases we
also show the measured properties of individual 2MASS clusters drawn from the
literature (Ponman et al. 1996; Ebeling \etal 1996, 1998; De Grandi et al. 1999; B\"ohringer \etal 2000; Mahdavi et al. 2000; Cruddace et al. 2002; Reiprich \& B\"ohringer 2002) and corrected to our assumed cosmology 
and to a bolometric X-ray luminosity.    

The $L_X-T$ relation of the stacked clusters closely resembles that derived from
observations of individual clusters (Fig.~\ref{fig:lt}).  There is a clear break in the slope between
high and low temperature clusters.  If we fit the results with a broken power 
law (see Table~\ref{tab:ltnfit}) we find slopes for rich ($2.7\pm0.7$) and poor 
($5.8\pm1.7$) clusters consistent with studies of individual clusters
($2.5-3.0$ for rich clusters, Wu et al. 1999; Rosati et al. 2002 
and $4-7$ for poor clusters, Helsdon \& Ponman 2000; Xue \& Wu 2000).  The correlation we observe
does not depend on the redshift when we subdivide our cluster catalogs into redshift bins.  
Similarly, the relationship between mass and temperature, shown in Fig.~\ref{fig:mt} for 
both $M_{200}$ and $M_{*666}$, is well-fit by a single power law with $M_{*666} \propto T^{1.79\pm0.17}$
and $M_{200} \propto T^{1.59\pm0.17}$ (see Table~\ref{tab:ltnfit}) The results are generally consistent
with relations derived from samples of clusters (e.g. Xu et al. 2001; Neumann \& Arnaud 1999; Sanderson et al. 2003; Arnaud et al.\ 2005b)
as well as with theoretical models (Evrard et al. 1996). The offsets between $M_{*666}$ and $M_{200}$ 
created by the differences between $r_{*666}$ and $r_{200}$ discussed in \S\ref{sec:cmass} have little 
effect on the results.  Finally, our estimated entropies are inconsistent with a simple
$S\propto T$ scaling relation, as shown in Fig.~\ref{fig:en}.  They are, however, consistent with
the shallower $S \propto T^{0.65}$ relation of Ponman et al. (2003) or a break in the entropy for
halos with the temperatures of order 1~keV where we observe a break in the $L_X-T$ relation.
In summary, the X-ray properties of the stacked clusters are essentially indistinguishable from
those found from studies of samples of individual clusters.

Figs.~\ref{fig:ln666}--\ref{fig:mn666} show the correlations of the X-ray properties and the mass
with the richness $N_{*666}$.  At a first glance, these correlations may seem to be inconsistent
with the X-ray and mass correlations we just discussed if the richness is a simple power-law
of the mass (Eqn. 4).  We carried out extensive tests to
verify these correlations such as subdividing the sample by redshift,
and using the soft-band rather than the bolometric luminosity.  Using the soft-band luminosity
avoids the need to accurately measure the temperature, and we expect any systematic problems
in the richness estimate to be strongly redshift dependent.  In fact, these correlations are the 
consequence of averaging at fixed optical richness in the presence of Poisson fluctuations. 
We can illustrate this by fitting all the correlations using 
the simple Poisson model of \S3.  We assume a power law relation for $T(M)$ and $N_{*666}(M)$,
a broken power-law relation for $L_X(T)$, and a cluster mass function cutoff at mass
scale $M_0$ over a mass range set by $\Delta$.   Most of these parameters are simply
the relations we need to fit the individual relations, with two parameters added to 
truncate the distribution of clusters at low mass.  The parameters of these relations
are summarized in Table~\ref{tab:ltnfit}, and we find a sensible mass cutoff of
$\log M_0/M_\odot \simeq 13.46 \pm 0.08$ that needs to be slightly softened by 
$\Delta \simeq  0.18 \pm 0.03$ in order to reproduce the gradual decline in the
X-ray temperatures and luminosities toward very low richness.  The mass-richness
relation is 
\begin{equation}
  \log N_{*666} = (1.10\pm0.04)+(0.87\pm0.05)\log(M/M_1).  
\end{equation}
Our earlier result
in Kochanek et al. (2003) was that $\log N_{*666} = (1.44\pm0.17)+(1.10\pm0.09)\log(M/10^{15}M_\odot)$ 
so the normalizations are consistent while the estimates of the slope differ by about
$2\sigma$.  The present, shallower slope is very similar to that of Lin, Mohr 
\& Stanford (2004) or Popesso et al. (2005).
Although the Poisson model is our preferred model, we also list the individual fits between stacked cluster 
properties in Table~\ref{tab:ltnfit} for comparison with other studies.

\begin{deluxetable}{ccccccccc}
\tabletypesize{\scriptsize}
\tablecolumns{9}
\tablewidth{0pt}
\tablecaption{Fitting Results for $L_X$--$T_X$--$N_{*666}$--$M$.\label{tab:ltnfit}}
\tablehead{
\colhead{Relation} &
\colhead{Reference} &
\colhead{Model} &
\colhead{$\log{Y_0}$} &
\colhead{$k$} &
\colhead{$\log{X_0}$} &
\colhead{$m$} &
\colhead{$n$} &
\colhead{$\chi^{2}_{\nu}(dof)$} 
}
\startdata
\cutinhead{Individual Fits\tablenotemark{a}}
$L_X-T$       & this paper & BPL & $43.11\pm0.23$ & \nodata\ & $0.13^{+0.59}_{-0.11}$ & $2.7\pm0.7$ & $5.8\pm1.7$ & 0.2(5) \\
$T-L_X$       & this paper & BPL & $0.10\pm0.04$  & \nodata\ & $43.02\pm0.7$ & $0.37\pm0.07$ & $0.15\pm0.07$ & 0.2(5) \\ 
$N_{*666}-L_X$ & this paper & SPL & $0.43\pm0.03$  &  $0.63\pm0.04$ & $43$ & \nodata & \nodata & 0.5(7) \\
               & K03\tablenotemark{b} & SPL & $0.31\pm0.06$  &  $0.75\pm0.05$ & $43$ & \nodata & \nodata & \nodata \\
$L_X-N_{*666}$ & this paper & SPL & $42.32\pm0.07$ & $1.56\pm0.11$ & $0$ & \nodata & \nodata & 0.5(7) \\
$N_{*666}-T$   & this paper & SPL & $0.04\pm0.04$ & $2.26\pm0.32$ & $0$ & \nodata & \nodata & 1.5(7) \\
               & K03        & SPL & $-0.25\pm0.14$ & $2.09\pm0.17$ & $0$ & \nodata & \nodata & \nodata \\
               & this paper & BPL & $0.56\pm0.13$  & \nodata & $0.13\pm0.13$ & $1.35\pm0.42$ & $4.0\pm1.2$ & 0.4(5) \\
$T-N_{*666}$   & this paper & SPL & $-0.01\pm0.03$ & $0.36\pm0.06$ & $0$ & \nodata & \nodata & 1.4(7) \\
               & this paper & BPL & $0.09\pm0.04$  & \nodata & $0.51\pm0.27$ & $0.74\pm0.17$ & $0.20\pm0.07$ & 0.4(5)\\ 
$M_{*666}-T$   & this paper & SPL & $13.64\pm0.05$ & $1.79\pm0.17$ & $0$ & \nodata & \nodata & 0.1(7) \\
$M_{200}-T$   & this paper & SPL & $13.56\pm0.05$ & $1.59\pm0.17$ & $0$ & \nodata & \nodata & 1.0(7) \\
$M_{*666}-N{*666}$ & this paper & SPL & $13.64\pm0.06$ & $0.74\pm0.10$ & $0$ & \nodata & \nodata & 1.1(7) \\ 
              & this paper & BPL & $13.79\pm0.09$ & \nodata & $0.48\pm0.30$ & $1.27\pm0.28$ & $0.36\pm0.17$ & 0.1(5)\\ 
$M_{200}-N{*666}$ & this paper & SPL & $13.58\pm0.06$ & $0.56\pm0.10$ & $0$ & \nodata & \nodata & 2.9(7) \\ 
              & this paper & BPL & $13.53\pm0.09$ & \nodata & $0.49\pm0.18$ & $1.44\pm0.27$ & $-0.05\pm0.17$ & 0.1(5)\\ 
\cutinhead{Poisson Model Fits\tablenotemark{c}}
$L_X-T$       & this paper & BPL & $42.55\pm0.05$ & \nodata\ & $0.00\pm0.05$ & $3.03\pm0.17$ & $10.3\pm2.4$ & \nodata \\
$M_{200}-T$   & this paper & SPL & $13.58\pm0.05$ & $1.65\pm0.12$ & $0$ & \nodata & \nodata & \nodata \\
\enddata

\tablecomments{The single power law (SPL) is defined by $Y = Y_0(X/X_0)^{k}$ while the broken power law (BPL) is 
 $Y = Y_0 (X/X_0)^m$ for $X> X_0$ and $Y = Y_0 (X/X_0)^{n}$
  for $X < X_0$. 
}
\tablenotetext{a} {Fits to scaling relations between two cluster parameters only.}
\tablenotetext{b} {Kochanek et al. (2003).}
\tablenotetext{c} {The fitting results for the global Poisson model (see \S3).}
\end{deluxetable}

\begin{figure}
\epsscale{0.7}
\plotone{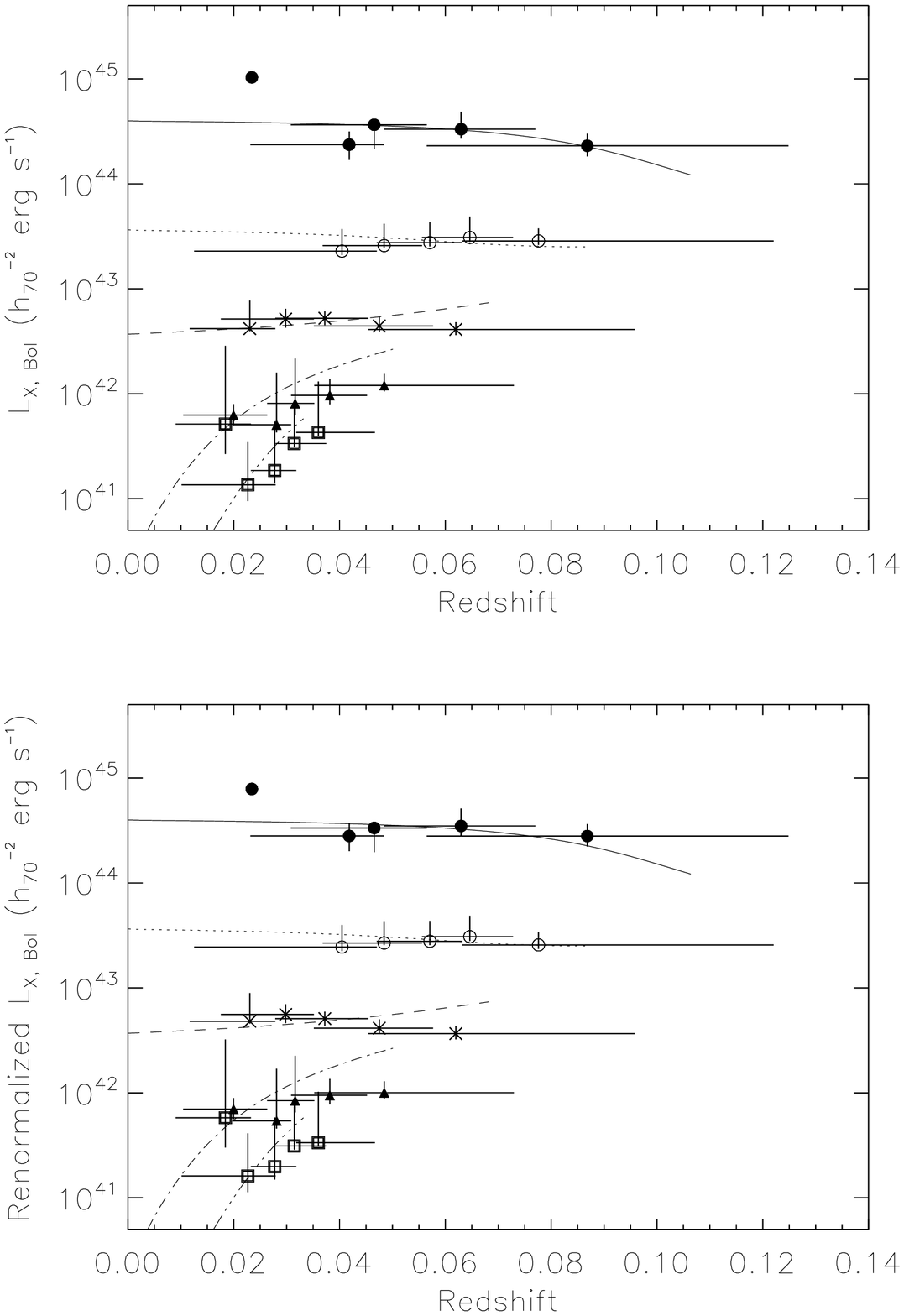}
\caption{X-ray luminosity versus redshift for stacked clusters in richness bins 0 (top) to 4 (bottom).  The 
  upper panel shows the raw results, the lower panel corrects for the difference between the mean richness
  of all clusters in the richness bin and the mean richness in the individual redshift bins.  The even and
  odd points in the redshift sequences are not independent measurements.
  The lines in both panels are the best fit Poisson models for each richness bin.
  \label{fig:lz}}
\end{figure}

\begin{figure}
\epsscale{0.7}
\plotone{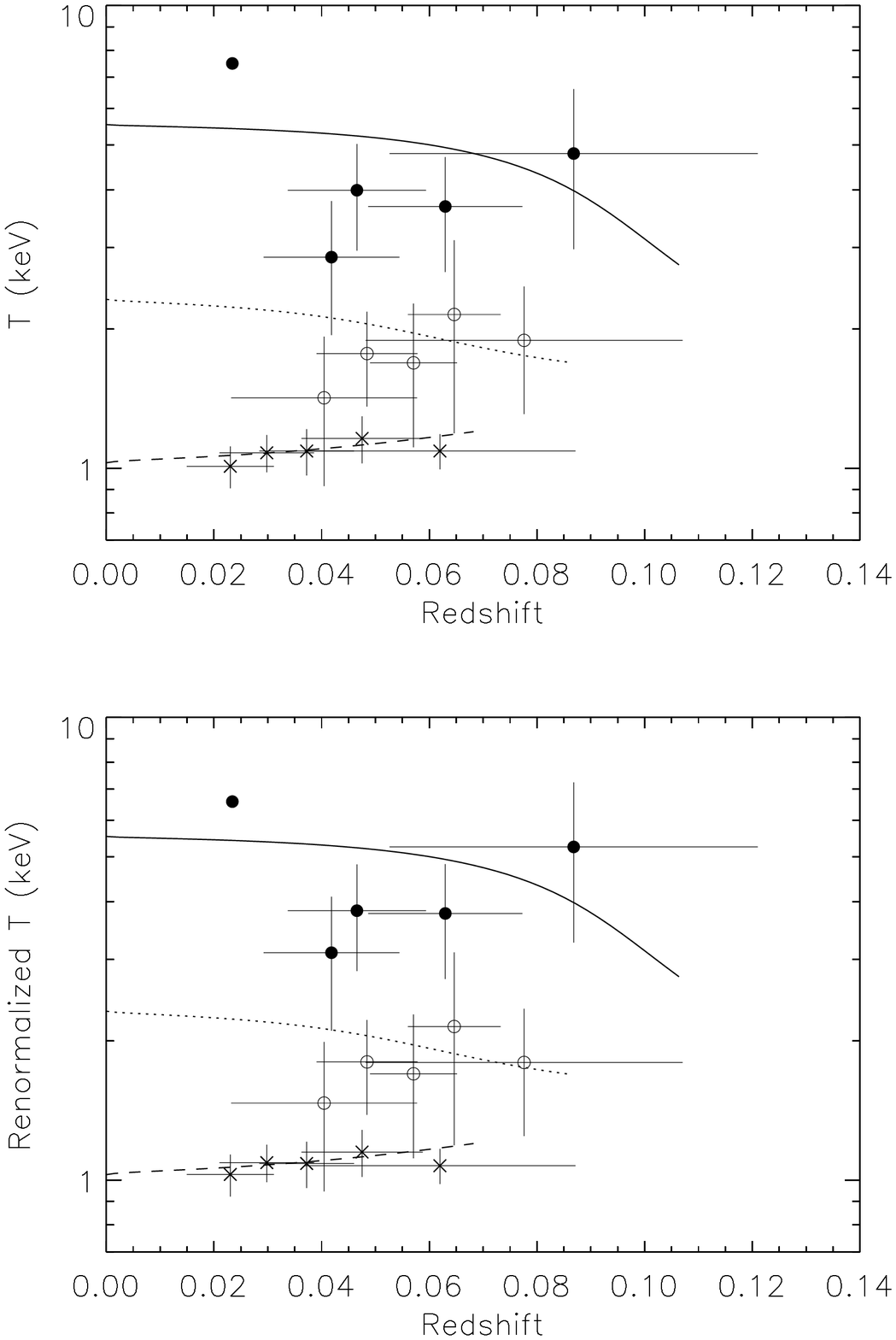}
\caption{Temperature versus redshift for stacked clusters in richness bins 0 (top), 1, and 2 (bottom).  The 
  upper panel shows the raw results, the lower panel corrects for the difference between the mean richness
  of all clusters in the richness bin and the mean richness in the individual redshift bins.  The even and
  odd points in the redshift sequences are not independent measurements.
  The lines in both panels are the best fit Poisson models for each richness bin.
  \label{fig:tz}}
\end{figure}

The last point we consider is whether the matched filter algorithm leads to cluster catalogs with redshift-dependent
biases.  We test for the biases by dividing each richness bin into three redshift bins with the boundaries chosen
so that each bin produced a stacked spectrum with a similar signal-to-noise ratio.  We also created two
overlapping (i.e. not independent) redshift bins centered on the edges of the first sets of bins to better illustrate
any variations with redshift. For each of these bins we extracted the spectra as described in \S4, normalized
each spectrum to the mean redshift of the bin, fit the spectra to obtain the luminosity and temperature and estimated the uncertainties 
using bootstrap re-sampling.  The results are shown in Figures~\ref{fig:lz} and \ref{fig:tz}, either as the
raw results or normalized to remove the variations in the mean richness between the redshift bins.  The error-bars
for temperatures are larger than those for the luminosities and we only show results from the first three richness bins.

The qualitative result is that there are few signs of redshift dependent biases beyond the expectations
of the Poisson model. The most massive clusters (richnesses 0 and 1) show a slow decline in the mean
luminosity and temperature at fixed richness because of the increase in the Malmquist biases with the
diminishing numbers of detectable galaxies.  The intermediate richness class 1 is far enough from both
the break in the mass function and the lower mass limit to show little variation with redshift.  The two lowest richness
classes show a rapidly rising luminosity with redshift.  In all cases, the trends are well
matched by the simple Poisson model -- we see no evidence for additional systematic biases.  

\section{Summary}

Optically-selected cluster catalogs have always been suspect because of concerns about false detections and
redshift-dependent biases created by chance alignments of galaxies.  Current attempts to understand the biases 
of optically-selected catalogs have focused either on testing the algorithms in mock galaxy catalogs 
(Kochanek et al. 2003; Miller et al. 2005) or by using weak lensing measurements (Sheldon et al. 2001).  
Unfortunately, there has never been a complete X-ray survey of a large optically-selected cluster catalog 
to provide an independent test of these concerns. In this study we have used a stacking analysis of the 
\rosat\ All-Sky Survey data to determine the average X-ray properties
of a sample of 4333 clusters found using the matched filter algorithm of Kochanek et al. (2003) applied to
the 2MASS galaxy survey with a magnitude limit of K$<13.25$~mag.  

After dividing the clusters into bins of optical richness, we could successfully measure the surface brightness
profiles, bolometric X-ray luminosities, X-ray temperatures, masses and gas entropies for the averaged clusters,
finding results that are generally consistent with those from analyzing individual clusters.  The one 
exception is that we tend to find larger core radii for $\beta$-model fits to the surface brightness
presumably because of additional smoothing created by errors in the optical positions.  The stacked
clusters cleanly follow the correlations observed for samples of individual clusters between X-ray
luminosity, temperature and mass.  However, in order to interpret the correlations between the optical
richness and the X-ray luminosity, temperature and mass, it is necessary to model the effects of
Poisson fluctuations of the number of galaxies in a cluster because in the presence of these
fluctuations averaging at fixed richness is very different from averaging at fixed mass. 
These effects matter for high richness clusters observed at redshifts where they will contain few 
detectable galaxies, and for all low richness clusters.  A very simple Poisson model for the effects of
these fluctuations naturally reproduces our results.  We note that Stanek et al. (2006) have
recently demonstrated similar effects arising from the scatter in X-ray luminosity at fixed
cluster mass.

We also examined the problem of redshift-dependent biases in optical catalogs by examining the 
X-ray properties of clusters of fixed richness as a function of redshift.  The low redshifts of the 2MASS clusters
($z<0.1$) means that we need worry little about genuine cosmological evolution.  Our statistical
model predicts that high richness clusters will show a slowly declining X-ray luminosity, while
low richness clusters will show a rising X-ray luminosity, and this agrees with our measurements.
Essentially, Malmquist biases created by the much more abundant low-mass clusters become 
steadily more important for the high mass clusters as the numbers of detectable member galaxies
decline, while the reverse occurs for the low-mass clusters.  In general, however, samples
restricted to clusters containing at least 3 detectable galaxies (already a floor required
to minimize the abundance of false positives) and to the higher mass clusters rather than
groups ($T \gs 2$~keV, $M \gs 10^{14}M_\odot$, $L_X \gs 10^{43}$~ergs/sec) have 
corrections due to the effects of Poisson fluctuations that are relatively simple to model
and depend little on the model for the cutoff between groups and galaxies.  Models for
the lower mass groups are more challenging, particularly if variations with redshift must
be included.  We see no evidence for large systematic errors created by chance superpositions.
 
Despite the shallowness of the RASS, our approach (see also Bartelmann \& White 2003) can be applied to 
analyses of cluster samples with
much higher mean redshifts than the 2MASS sample, where even the richest clusters are tracked only
to $z \ls 0.1$.  Modulo redshift scalings for K-corrections and between cosmological distances,
the signal-to-noise ratio of a stacking analysis scales as the square root of the product of the survey
area and depth.  Thus, an analysis of clusters in 10000~deg$^2$ of the SDSS survey (25\% of the
area of 2MASS), which can detect rich clusters with $z\ls 0.5$ rather than $z\ls 0.1$ 
(5 times the depth), should have a signal-to-noise ratio comparable to our present
analysis for the measurement of the mean X-ray properties of the clusters.  Analyses of rich
clusters over 1000~deg$^2$ would have a signal-to-noise ratio comparable to our analysis of
clusters with $\sim 1/3$ the richness.  This assumes that the analysis remains limited by 
statistical errors rather than being dominated by systematic problems such as contamination by
faint point sources or difficulty in controlling the background.

\acknowledgements
This publication makes use of data products from the Two Micron All Sky Survey, which is a joint project of the University of Massachusetts and the Infrared Processing and Analysis Center/California Institute of Technology, funded by the National Aeronautics and Space Administration and the National Science Foundation.
This work makes use of the RASS data achieved by the HEASARC, a service of the Exploration of the Universe Division at NASA/GSFC
and the High Energy Astrophysics Division of the Smithsonian Astrophysical Observatory.  We thank J. Huchra for his assistance
in generating the new group catalogs.  We thank M. F. Corcoran, S. Snowden,  
D. Grupe, E. Rykoff and T. Mckay for helpful discussion on the stacking of RASS data, and M. W. Bautz, G. Evrard, D. Weinberg, M. White, Z. Zheng for 
their comments on the results.


\end{document}